\documentclass[conference]{IEEEtran}
\IEEEoverridecommandlockouts
\usepackage{balance}
\usepackage{cite}
\usepackage{amsmath,amssymb,amsfonts}
\usepackage{algorithmic}
\usepackage{graphicx}
\usepackage{textcomp}
\usepackage{xcolor}
\usepackage{diagbox}
\usepackage[latin1]{inputenc}
\usepackage{tikz}
\usetikzlibrary{shapes,arrows}
\title{
Anchoring the Value of Cryptocurrency 
}
\begin{document}

\author{\centering
\IEEEauthorblockN{Yibin Xu}
\IEEEauthorblockA{\textit{School of Computer Science}\\\textit{ and Informatics} \\
\textit{Cardiff University}\\
Cardiff, UK \\
work@xuyibin.top}
\and
\IEEEauthorblockN{Yangyu Huang}
\IEEEauthorblockA{\textit{School of Electronic Engineering}\\\textit{ and Automation} \\
\textit{Guilin University of Electronic Technology}\\
Guilin, China \\
i@hyy0591.me}
\and
\IEEEauthorblockN{Jianhua Shao}
\IEEEauthorblockA{\textit{School of Computer Science}\\\textit{ and Informatics} \\
\textit{Cardiff University}\\
Cardiff, UK \\
shaoj@cardiff.ac.uk}
}
\maketitle
\begin{abstract}
A decade long thrive of cryptocurrency has shown its potential as a source of alternative-finance and the security and the robustness of the underpinning blockchain technology.

However, most cryptocurrencies fail to show inimitability and their meanings in the real world. As a result, they usually start off as favourites but quickly become the outcasts of the digital asset market.

The blockchain society attempts to anchor the value of cryptocurrency with real values by employing smart contracts and link it with computation resources and the digital-productivity that have value and demands in the real world. But their attempts have some undesirable effects due to a limited number of practical applications. This limitation is caused by the dilemma between high performance and decentralisation (universal joinability). The emerging of blockchain sharding models, however, has offered a possible solution to address this dilemma.

In this paper, we explore a financial model for blockchain sharding that will build an active link between the value of cryptocurrency and computation resources as well as the market and labour behaviours. Our model can adjust the price of resources and the compensation for maintaining a system based on those behaviours. We anchor the value of cryptocurrency by the amount of computation resources participated in and give the cryptocurrency a meaning as the exchange between computation resources globally. Finally, we present a working example which, through financial regularities, regulates the behaviour of anonymous participants, also incents/discourages participation dynamically.
\end{abstract}


\section{Introduction}
Money is used universally, representing a medium of exchange, a unit of account, a store of value, and a standard of deferred payment \cite{murad1943nature}. Currency is always associated with national identity, safeguarded by the sovereignty, required to be inimitabe and used in daily life by a population \cite{aschheim2006money}. Currency is also a tool and voice in international politics \cite{goldberg2008vehicle,rey2001international,kamps2006euro} - a currency is usually priced by the national power and nonrenewable resources that a country holds \cite{goldberg2008vehicle,kamps2006euro}. Centralised currency policies can incent and exploit labour and productivity, and have coupling effects on market behaviour \cite{gros1996national,maarek2013currency}. 

When considering a cryptocurrency, people usually have the impression that it has loose regulation, high privacy and fairness \cite{narayanan2016bitcoin,hileman2017global,fry2016negative}. But people cannot associate the value of a cryptocurrency with any sovereign or nonrenewable resources like they do for a normal currency \cite{yermack2015bitcoin,ali2014economics,bouoiyour2016drives}. That is, cryptocurrencies do not have a value anchor, are less connected to our daily life, and are not difficult to build or replace in the current means of human economics.

In technical terms, every transaction in a cryptocurrency needs to be broadcast to avoid the double-spend problem in a decentralised environment \cite{nakamoto2019bitcoin}. However, ordinary people who hold home desktops, laptops or mobile devices cannot be expected to do this constantly, nor be able to be fully synchronised with the environment most of the times \cite{semyonova2017past,gencer2018decentralization}. That is, for most people, their devices will not in be in the system all the time and only a very small number of devices compared to the user population will actually act as maintainers\cite{semyonova2017past}. These phenomena make cryptocurrencies unreliable and easily abandoned\cite{bartoletti2020dissecting,urquhart2016inefficiency,eyal2018majority,raymaekers2015Cryptocurrency}.

Ethereum \cite{luu2016making} and other smart contract approaches \cite{bhargavan2016formal,luu2016making,wood2014ethereum} try to form a more decentralised sovereign where ``citizens" exchange productivity through smart contracts. However, this approach tends to result in few citizens actually doing jobs on the platform, and because blockchains require all citizens to witness and conduct every job to ensure the integrity of results by consensus, productivity is reduced. Some citizens may be disadvantaged, hence eliminated, if the system does highly sophisticated jobs or an overloaded workflow. Consequently, the performance of these platforms will stay low in order to attract citizens, making smart contract platforms not even being able to perform simple tasks such as powering a game of digital pet smoothly \cite{perezanalysis}.

With the idea of blockchain sharding \cite{dang2019towards,zamani2018rapidchain,kokoris2018omniledger} which assigns different citizens to do different tasks in parallel securely, it is promising that productivity can be improved and the usage of cryptocurrency can be extended to, for example, exchanging computation power for doing more sophisticated jobs and increasing performance of computation resources. Because transactions are divided into shards, the requirement on citizens' computation ability and network bandwidth is much reduced due to reduced workflows, more citizens can be added to this type of decentralised sovereign, and it makes it easier for data synchronisation. However, digital citizens are different from citizens in the real world. Just like characters in a video game, digital citizens can die and then start again, and can ignore regulations causing damages without punishments \cite{yasaweerasinghelage2017predicting,moinet2017blockchain}. Since computation resources are utilised to fulfill the demand in the real world, the cost of using such resources has meaning in the real world too\cite {freund2018economic}. With the smart contract approach, cryptocurrency money is earned with real effort and real labour which has a price in reality. Therefore, it should be possible to regulate digital citizens by, for example, confiscating their digital assets if they behave inappropriately, or encouraging citizens to perform jobs by exciting them with compensation. In this way, it is possible for the system to stabilise the performance of participated resources without knowing the source of them in reality or building a credit base identity.

Grid Computing \cite{joseph2004grid} and Cloud Computing\cite{hayes2008cloud} are two computational models that utilise resources globally through the Internet. Grid computing attempts to use the idle resources globally and is a multi aid platform with credit-based identity model. Cloud computing, on the other hand, is a model where users buy resource globally owned and priced by a centralised organisation. We see the popularity of cloud computing in both industry and academia \cite{foster2008cloud,buyya2008market,rittinghouse2017cloud}. Cloud platforms have generated considerable revenues \cite{au2016industry}, and they have changed the way technology industry operates, including the way to host a website or buy computation resources. In contrast, Grid computing is less popular, due to these main difficulties: 
(1) needs necessary interfaces for data exchange and job execution on different devices and using different software to complete a task. 
(2) needs to be able to trust job results returned from an anonymous source.
(3) needs to regulate participated devices worldwide, without having to know the identity of the owners, and to ensure jobs are delivered correctly and on time.

We believe that blockchains and cryptocurrencies can be exploited to address some of the problems of Grid computing we outlined here. The first challenge can be addressed by the use of smart contracts, which can unify interface of data exchange and job executions. The second issue can be tackled by blockchain, especially, blockchain sharding techniques, where the consensus job results returned from a shard is guaranteed to be correct as long as the security threshold is maintained. Compensations and punishments in a cryptocurrency associated with the blockchain can be employed to address the quality of return. By addressing these issues with blockchain sharding and smart contract techniques, the Grid computing techniques may be improved. However, to function a Grid computing platform and to make it a decentralised sovereign, economic regulation and policy must serve as a tool to adjust and control the balance between digital labour and demand in the market. The problems of inflation, deflation or even economic break down that exist in real world economy can also exist in the digital world. Therefore, an economic model in this type of decentralised sovereign is important to study. 

In the remaining of the paper, we will first introduce blockchain, smart contract and blockchain sharding. Then, we discuss an autonomous finance model that use the classical money supply indicators of M0, M1, M2, and the quantity theory of money \cite{mccandless1995some,friedman1989quantity} to price a currency by service demand and to encourage and discourage labour participation by wages. We use linear regression\cite{seber2012linear} to set the parameters to regulate the market and we provide a simulated experiment at the end of the paper.

\section{Preliminaries}

\subsection{Blockchain}
Proof of Work (PoW) describes a system that is difficult to be created but easy to be verified. The most widely used Proof-of-Work scheme - Hashcash\cite{back2002hashcash}, is based on SHA-256 and was later introduced as a part of Bitcoin (Nakamoto blockchain) as the computation strength competition method. There are different kinds of PoW alternatives proposed for blockchain \cite{vasin2014blackcoin,xu2019mwpow}.

A block in Nakamoto blockchain embeds the information of a period; the blockchain periodically attaches new blocks to the blockchain. The difficulty is a measure of how difficult it is to generate a PoW.
\begin{gather}Difficulty= \frac{Difficulty\_1\_target}{Current\_target}\end{gather} where $Difficulty\_1\_target$ is a constant 256 bit number and $Current\_target$ is a 256 bit number. When calculating the difficulty for a hash, the hash itself is used as the $Current\_target$, then $Difficulty$ can be derived by (1). The Nakamoto blockchain network has a global block difficulty: valid blocks must have a hash below the current target. The hash is adjusted by changing the value of Nonce (a field in the block). The global difficulty is adjusted to limit the rate at which the network can generate one new block in an approximately fixed time interval.

Nakamoto blockchain has a pre-defined security threshold: the honest people must take more than $50\%$ of computation power. This security threshold guarantees that the malicious people do not have enough power to create a longer fork branch of blocks when honest people are working on another branch. New participants can determine the correct records by staying with the longest chain (the mainchain) and we suppose this chain is longer than the second-longest chain for at least a given length.

No one should be able to send the same money to more than one receiver at the same time; this is the essential requirement for a decentralised-cryptocurrency. This requirement is fulfilled when participants can determine the correct record by checking whether the sender has spent the money or not in the history of the mainchain.
\subsection{Smart Contract}
A Smart Contract is a Turing Complete \cite {aman2009turing} computer protocol that is intended to digitally facilitate, verify, or enforce the negotiation or performance of a contract \cite{clack2016smart}. Smart contracts allow the performance of credible transactions without third parties \cite{clack2016smart,mccorry2017smart}. These transactions are trackable and irreversible. A smart contract can be used as a unified interface for job assignment and result verification. When executing a smart contract, the user transfer an amount of funding into the smart contract address, each code execution is priced by the number of lines of codes executed.

\subsection{Blockchain Sharding}
Blockchain sharding is an improvement to the basic blockchain design which aims to achieve a secure consensus together with a sub-group of people among the whole population. Because the job workload and the nodes are divided into shards, nodes only need to process the data in a Shard and mostly communicate with nodes in a shard. This significantly reduces the requirement on computation capability and bandwidth. Also, because the shards are running in parallel, the overall processing ability is also increases tremendously. The main challenge is to inhibit the chance for adversaries who do not control the majority people globally but hold the majority people in a sub-group to temper the record. To bound the maximum chance for an adversary to gain control of a sub-group under a given security threshold, the number of sub-groups (shards) the system can have and the number of people (node) in the sub-groups are strictly restricted. Also, the model must fulfill the following requirements:
(1) people cannot choose which subgroup they would be located in;
(2) a transparent and random node assignment scheme must be employed, meaning no one can predict or manipulate which shard it is about to be assigned in;
(3) the number of shards must be dynamically adjusted with the change of population.

We can calculate how many times of node assignments is required to guarantee an Adversary in controlling a Shard. The probability of obtaining no less than $x$ adversary nodes when randomly picking a shard sized $m$ ($m$ number of nodes inside the Shard) can be calculated by the cumulative hypergeometric distribution function without replacement from a population of $n$ nodes. Let $X$ denote the random variable corresponding to the number of adversary nodes in the sampled Shard and $t$ is the number of adversary. The failure probability for one committee is at most

\begin{gather}Pr[X>[m/2]]=\sum^m_{X=[m/2]}\frac{(^t_X)(^{n-t}_{m-X})}{(^n_m)} \label{eq:r2} \end{gather}

\begin{figure}[htbp]
	\begin{tabular}{c|c}
   \includegraphics[width=0.215\textwidth]{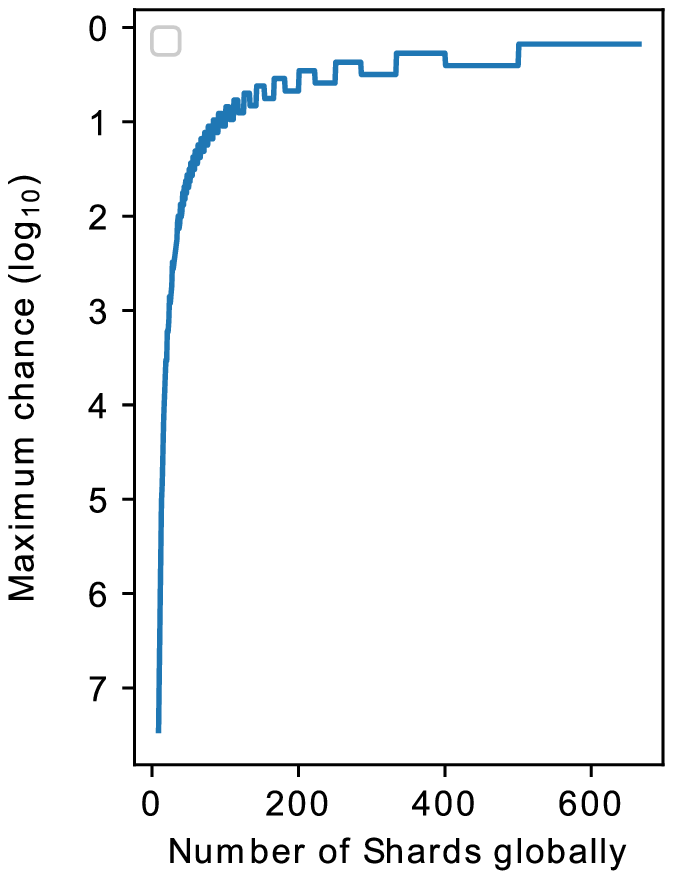}&\includegraphics[width=0.230\textwidth]{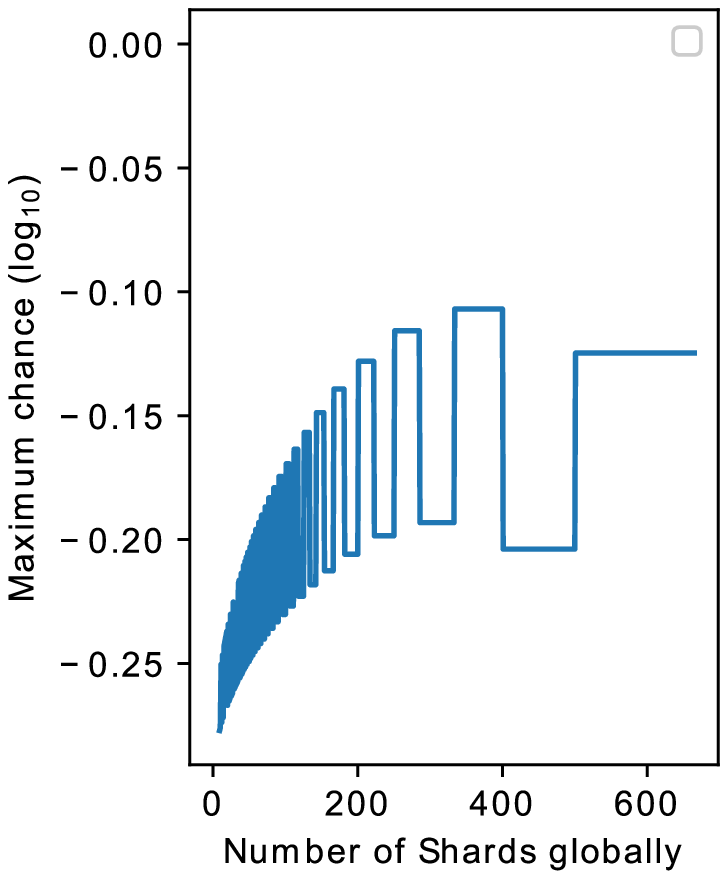}\\
		$t=n/3=666$&$t=n/2=1000$\\
\end{tabular}
	\caption{The chance to fail when $n=2000$, $t=n/3$, $t=n/2$ and $m=n/s$ where $s$ is the number of Shards}
	\label{fig:img2}
\end{figure}

Figure \ref{fig:img2} shows the maximum probability to fail with $n=2000$ and $m=n/s$ where $s$ is the number of Shards. As can be seen from the result, the system has a very high failure chance when the adversary taken $n/2$ of nodes. We expect blockchain sharding approaches to maintain the same $n/2$ security level as in the original blockchain systems, but as shown in figure \ref{fig:img2}, the most blockchain sharding approaches can only withstand up to $n/3$ of nodes being bad, and only a few Shard can exist.

\subsection{N/2 Adversary Resistant Blockchain Sharding}
We previous proposed a \emph{Jury} Hypothesis that serves as an analogy of an $n/2$ Byzantine-node tolerate blockchain sharding approach in \cite{XUXU}, which further improves blockchain sharding.

The \emph{Jury} hypothesis states that the member of a Jury of a court comes from the diverse background, so that when a verdict is reached, it can be seen as the decision was reached from the whole society (every class of people). If it takes $m$ different occupations to form a jury, then when there are a $s$ number of court hearings running in parallel, there are $s$ number of people in each one of the $m$ occupation. Table \ref{fig:img3} shows a court schedule; each court represents a Shard, $A$ is a person controlled by the Adversary while $H$ is an honest person.

\begin{table}[h!]
\centering
\caption {Court Jury Schedule}
\begin{tabular}{ccccc}
\diagbox{Ocp}{Court number}&0&1&2&3\\
Occupation 1&A&A&A&A\\
Occupation 2&H&A&H&A\\
Occupation 3&A&H&A&H\\
Occupation 4&H&A&H&H\\
Occupation 5&H&H&H&H\\
\end{tabular}
\label{fig:img3}
\end{table}

It is ruled that a verdict is reached when a pre-defined $T,\ T>0.5m$ number of people inside the jury reached a consensus. Assuming there exists a random assignment scheme that assigns people of the same occupation to different courtrooms where different court hearings are taken place in parallel. Then, the chance for the Adversary to gain $T$ spots inside the target courtroom is (assuming without loss of generality that the adversary puts all its nodes into the front $T$ occupations)
\begin{equation}
    Pr[T]=\prod_{i=1}^{T}\frac{A_i}{s}
\end{equation}
where $A_i$ is the number of people inside courtroom $i$ who are controlled by the adversary. To derive the maximised $Pr[T]$, we want $\prod_{i=1}^{T}A_i$ to be maximised because $s$ is the same. Let the Adversary has $AD$ number of people inside the system (Court Jury Schedule), then $AD=\sum_{i=1}^m A_i$. To maximise the value of $\prod_{i=1}^{T}A_i$, we consider 
\begin{equation}
    A_i=\lfloor AD/T\rfloor, i \in [1,T-1]
\end{equation}
\begin{equation}
    A_T=\lfloor AD/T\rfloor+ AD\ mod\ T
\end{equation}
This scenario is the maximised because, given any positive integer $X$, 
\begin{equation}
    X*X > (X-1)*(X+1)=X*X-1
\end{equation}

\begin{figure}[htbp]
	\begin{tabular}{l|l}
   \includegraphics[width=0.22\textwidth]{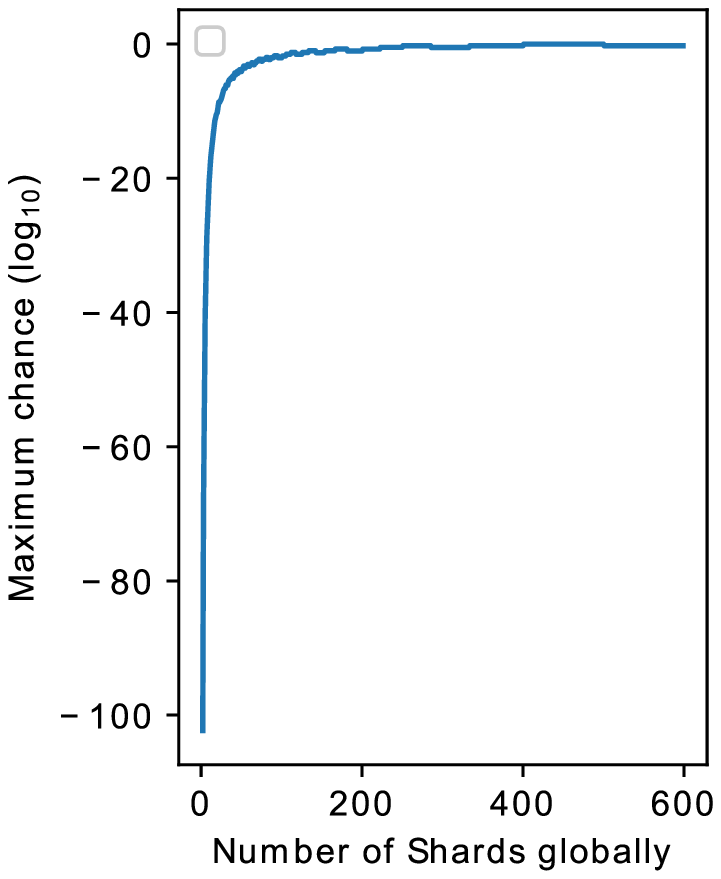}&\includegraphics[width=0.22\textwidth]{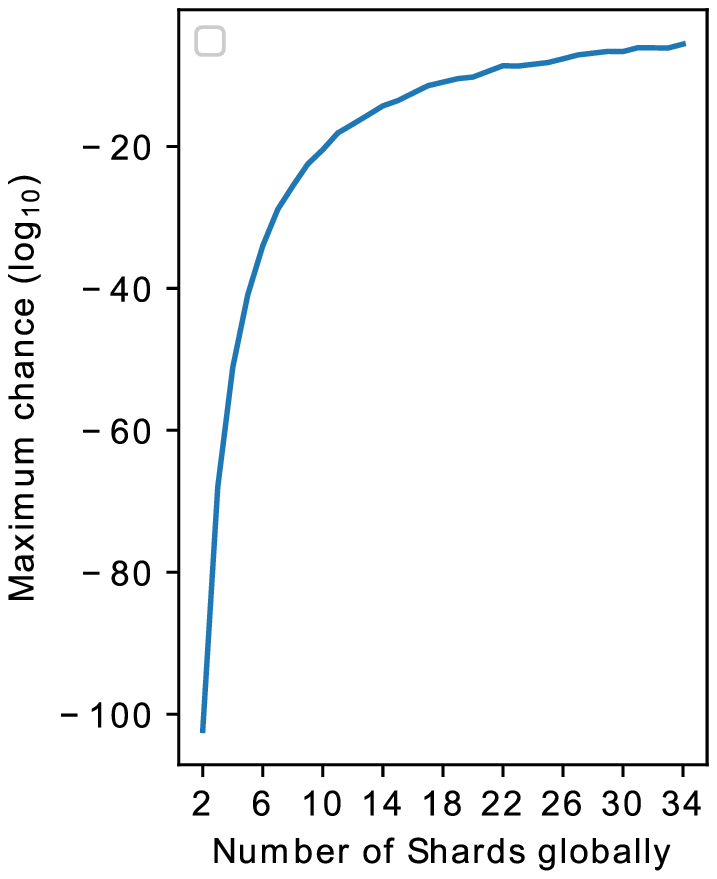}\\
		\small{$s \in [2,600], AD=1000=n/2$}&\small {$s \in [2,34], AD=1000=n/2$}\\
\end{tabular}
	\caption{The chance to fail with different $s$ when $n=2000$ and $m=n/s$ where $s$ is the number of Shards;}
	\label{fig:img4}
\end{figure}

Thus, 
\begin{equation}
    Pr[T]_{max}\approx(\frac{AD}{T*s})^T
\end{equation}
Though the adversary cannot manipulate a sentence when it does not have $T$ people inside a Shard, it can halt a sentence to be reached when it has $m-T+1$ number of the nodes in a Shard. Then this sentence cannot be made until the next court (the group of juries are re-selected). Thus, to make the system function more smoothly, we want $T\approx[m/2]$ while meeting the security threshold (e.g. $10^{-6}$ failure chance). Figure \ref{fig:img4} shows the maximum failure chance with different $s$, $n=s*m=2000$, $T=0.7*m$ and $AD=1000$ ($1/2$ fraction of the overall population).

As can be seen from the result, when there are ten Shards and $n/2$ people being evil, the failure chance is below $10^{-20}$, which significantly outperformed the traditional blockchain sharding at below $10^{-6}$ when it has ten Shards and only $n/3$ nodes being evil. If we set the block interval to be $30\ minutes$, then it takes over $10^{15}$ years to fail the system. If it is $10\ minutes$ (the same as Nakamoto blockchain), it still takes over $10^{14}$ years to fail. If we maintain a $10^{-6}$ failure chance at this circumstance with $T=0.7*m$, then there can be $33$ Shards at the same time.

In \cite{XUXU1}, there is a method that dynamically splits and combines the occupations to recover the system from a global halting (all shards are halted at the same time).

\section{Economic model}
\begin{figure}[h!]
\begin{tikzpicture}[auto,
    %decision/.style={diamond, draw=black, thick, fill=white,
    %text width=8em, text badly centered,
    %inner sep=1pt, font=\sffamily\small},
    block_center/.style ={rectangle, draw=black, thick, fill=white,
      text width=6em, text centered,
      minimum height=2em},
    block_left/.style ={rectangle, draw=black, thick, fill=white,
      text width=16em, text ragged, minimum height=4em, inner sep=6pt},
    block_noborder/.style ={rectangle, draw=none, thick, fill=none,
      text width=18em, text centered, minimum height=1em},
    block_assign/.style ={rectangle, draw=black, thick, fill=white,
      text width=18em, text ragged, minimum height=3em, inner sep=6pt},
    block_lost/.style ={rectangle, draw=black, thick, fill=white,
      text width=16em, text ragged, minimum height=3em, inner sep=6pt},
      line/.style ={draw, thick, -latex', shorten >=0pt}]
    % outlining the flowchart using the PGF/TikZ matrix funtion
    \matrix [column sep=2mm,row sep=5mm] {
      % enrollment - row 1
     \node[block_center, draw=none] (l){\small {Initial funding distribution}};&&&\node [block_center] (pool) {Funding pool};\\
     \node [block_center] (SM) {Smart Contract account}; &&&\node [block_center] (TA) {Transaction account}; &&&\node [block_center] (Margin) {Margin account};\\
      % follow-up - row 5
    };% end matrix
    % connecting nodes with paths
    \begin{scope}[every path/.style=line]
      % paths for enrollemnt rows
      \path (l)  -- (pool);

\path (pool)  -- (TA);
\path (TA)  -- (pool);
\path (SM)  -- (TA);
\path (TA)  -- (SM);
\path (Margin)  -- (TA);
\path (TA)  -- (Margin);
\path (Margin)  -- (pool);
\path (SM)  -- (pool);
    \end{scope}
  \end{tikzpicture}
\caption{Relationships between accounts and the funding pool}

\label{fig:1}
\end{figure}
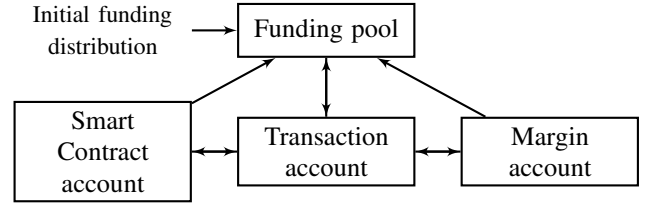

In this section, we show an economic model for a $n/2$ Blockchain sharding system with a halting recovery method. Smart contract is adopted in this blockchain sharding system. There are a funding pool and three types of accounts in our model.
\begin{enumerate}
    \item \emph{Transaction Account}. An address used for receiving compensation, keeping funding, and it can send and receive cryptocurrency to and from others. There is a fixed amount of transaction fee for every transaction related to the Transaction account. The fee is submitted to the funding pool automatically when the transaction is embedded into a block.
    \item \emph{Smart Contract Account}. A smart contract is associated with an account; every time the Smart Contract is executed, an amount of funding is sent to the system from this account. The funding of a Smart contract account initially comes from a Transaction account; the remaining funding can only be transferred back to that Transaction account.
    \item \emph{Margin Account}. The funding in a Margin account is frozen and cannot be used during a period.
\end{enumerate}
 Figure \ref{fig:1} shows the relationship between the accounts and the funding pool.

Grid computing models may run in mobile devices. It is common for mobile devices to go offline without prior notices due to the lousy network or low battery. This situation will not affect our model from functioning as long as there is a sufficient number of reliable nodes. We encourage reliable nodes to participate in the system by giving them compensation. In order to qualify for the compensation, the nodes need to first register as reliable nodes. They need to send an amount of funding from their \emph{Transaction account} into the \emph{Margin account} when registering. This funding will be unfrozen and being transferred back to the Transaction account if they stay online for a given period and carry out the duty by voting for consensus. The compensation is an additional fraction of the funding sent to the Margin account. The exact number of this additional fraction is changed by times (which is similar to the interests in time deposits in real life). If a reliable node goes offline within the period, the relevant funding in its Margin account is confiscated to the funding pool.

The price for executing smart contracts are changed time by time. This price is charged by the system, so that when a Smart Contract is executed, an amount of funding will be transferred from the smart contract account of this smart contract to the funding pool.

The same as Nakamoto blockchain, there is an initial funding distribution, where there is an amount of new funding released from the system in every block interval. This funding gradually reduces to zero block by block. After the initial funding distribution is finished, the money for executing the smart contract is the main source of funding for the funding pool to pay the compensation of the mining game as well as the compensation for reliable nodes.

\subsection{The Pricing Model}
We define the following financial indicators.
\begin{enumerate}
    \item $M_0$: The amount of currency in Transaction accounts.
    \item $M_1$: $M_1$ = $M_0$ + \emph{funding in Smart contract accounts}.
    \item $M_2$: $M_2$ = $M_1$ + \emph{funding in Margin accounts}.
    \item $Q$: The total lines of smart contract codes executed in a block interval.
    \item $P$: The price for executing a line code in smart contracts.
    \item $AVGQ$: The average of $Q$ over a long time window.
    \item $U$: The ideal coefficient of the $M_2$ used for executing smart contracts in a block interval.
\end{enumerate}
P is adjusted in every block interval.
\begin{equation} 
P_X=\frac{U\times M_{2_{X-1}}}{AVGQ+1}
\label{eee}
\end{equation}
where $P_X$, $M_{2_{X-1}}$ are the $P$ and $M_2$ at the block interval $X$ and $X-1$, respectively.

\subsection{The Rewarding Model}
We define the following three indexes.
\begin{enumerate}
    \item $GPL_X$: The required length of time for a new reliable node at block interval $X$ to stay online and carry out the duties. When a node transfers funding to its Margin account, it will become a reliable node after this transaction is embedded to a block.\footnote{a node can be multi-reliable if it transfers funding to the Margin account when it is already a reliable node. The potential confiscation of funding and the compensation for participating are being count separately.}
    \item $GN_X$: The number of new reliable nodes at the beginning of the block interval $X$.
    \item $I$: An indicator of the amount of funding in total to pay the compensation of nodes which start being reliable nodes at the beginning of the block interval $X$.  
\end{enumerate}

Let $R_X$ be the amount of funding flood to the funding pool in total during the $X-1$ block interval. The funding consists of the transaction fees, the fees for executing the smart contracts and the money from the initial currency distribution.

In every block iteration, $I\times R_X$ amount of funding is remained in the funding pool and will be used to pay the nodes who start to serve as reliable nodes at block height $X$ and finish serving at the block height $X+GPL_X$. Assumed an unreliable node $Alice$ starts to be reliable at block height $X$, the amount of compensation she will get at the block height $X+GPL_X$ is 
\begin{equation} \frac{Margin_{Alice}}{Margin_{\{X\}}}\times I\times R_X
\end{equation} 
where $Margin_{Alice}$ is the funding Alice frozen and $Margin_{\{X\}}$ is the total amount of funding that new reliable nodes at the block interval $X$ frozen. 

The "citizens" who participated in maintaining the system (propose blocks and/or vote for consensus) in block interval $X-1$ will equally divide the remaining $R_X$ at block height $X$.
\subsection{Economic Policies}
We define $B$ as a fixed number of $\frac{M_2}{M_1}$.

We change the economic policies to make $\frac{M_2}{M_1} \approx B$ in every block interval. If $B=2$, meaning, ideally, $50\%$ of the overall currency should be placed in Margin accounts.

We use the L2 Regularisation \cite{moore2011l1} to analysis $GPL$,$GN$, $I$ and $\frac{M_{2}}{M_{1}}$. Let
\begin{equation}
f(GPL_X,GN_X,I_X,\frac{M_{2_X}}{M_{1_X}})=abs(B-\frac{M_{2_{X+1}}}{{M_{1_{X+1}}}})\\
\end{equation} where $X \in [CH-100, CH-1)$, $CH$ is the current block height. 

We can then get the predicted $GPL_{CH}$,and $I_{CH}$ of $$f_{min} (GPL_{CH},GN_{CH},I_{CH},\frac{M_{2_{CH}}}{M_{1_{CH}}})$$ The predicted $GPL_{CH}$ and $I_{CH}$ then become the $GPL$ and $I$ at the block height $CH$.

Because the $GPL_X, GH_X, I_X, X\in[CH-100,CH-1)$ are data that every node in the system knows, every node can calculate the same $GPL$ and $I$ at the block height $CH$. So that, the economic policies are derived by the pre-defined rules instead of decided by some centralised authorities.
\subsection{Interaction Between Currency and Resources}
The maximum performance (transactions per second) of a blockchain sharding system is bound by the number of the Shards and the pre-defined maximum performance per shard. Thus, $Q$ has upper limits. When every Shard is fully loaded, many unconfirmed transactions and smart contracts will be left in pending status. The funding associated with those pending transactions and smart contracts will also stay in pending status. Then, that affects the mobility of funding and decreases $\frac{M_2}{M_1}$. 

Assuming at a block iteration, $U'\times M_2$ is the amount of money used to purchase resources, and $AVQ$ is equal to the $Q$ at its upper limit (fully loaded). According to equation \ref{eee}, when $U'>U$, $P$ goes up, and vice versa.

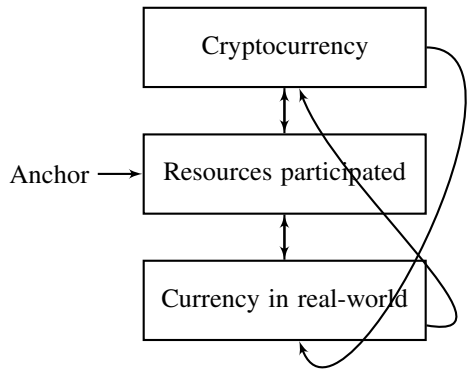
\begin{figure}[h!]
\centering
\begin{tikzpicture}[auto,
    %decision/.style={diamond, draw=black, thick, fill=white,
    %text width=8em, text badly centered,
    %inner sep=1pt, font=\sffamily\small},
    block_center/.style ={rectangle, draw=black, thick, fill=white,
      text width=10em, text centered,
      minimum height=3em},
    block_left/.style ={rectangle, draw=black, thick, fill=white,
      text width=16em, text ragged, minimum height=7em, inner sep=6pt},
    block_noborder/.style ={rectangle, draw=none, thick, fill=none,
      text width=18em, text centered, minimum height=1em},
    block_assign/.style ={rectangle, draw=black, thick, fill=white,
      text width=18em, text ragged, minimum height=7em, inner sep=6pt},
    block_lost/.style ={rectangle, draw=black, thick, fill=white,
      text width=16em, text ragged, minimum height=7em, inner sep=6pt},
      line/.style ={draw, thick, -latex', shorten >=0pt}]
    % outlining the flowchart using the PGF/TikZ matrix funtion
    \matrix [column sep=2mm,row sep=6mm] {
      % enrollment - row 1
     &&&\node [block_center] (c) {Cryptocurrency};\\
     \node (A) {Anchor};&&&\node [block_center] (r) {Resources participated}; \\
     &&&\node [block_center] (M) {Currency in real-world};\\
      % follow-up - row 5
    };% end matrix
    % connecting nodes with paths
    \begin{scope}[every path/.style=line]
      % paths for enrollemnt rows
\path (c)  -- (r);
\path (r)  -- (c);
\path (r)  -- (M);
\path (M)  -- (r);
\path (A)  -- (r);
\draw    (c) to [out=0,in=-70] (M);
\draw    (M) to [out=-10,in=-70] (c);
    \end{scope}
  \end{tikzpicture}
\caption{The relationship between Cryptocurrency, resources and real-world currency}

\label{fig:23}
\end{figure}

$P$ affects the purchase demands, as well as the funding for the rewards. When $P$ go up, the amount of reward is also increased, this can drive more devices into participating in the system. When $P$ go down, the amount of reward is decreased, this surpasses the devices from participation. With the more devices participated, the more Shards can be formed, and that changes the upper limit of $Q$, vice versa. By the appropriate adjustments of the parameters, we can anchor a cryptocurrency with demands from the purchase market as well as the availability of the labour market. There is a derivation relationship between the exchange of the cryptocurrency and the resources and the exchange of the resources and real-world costs (electric bills and other fees to maintain the computation resources). Thus, the real-world value is placed to this cryptocurrency by the computation resources serving as the middle man. Figure \ref{fig:23} shows the relationship between the cryptocurrency, computation resources and the currency in real world.

\section{Experiment}
\begin{figure}[htbp]
	\begin{tabular}{l}
   \includegraphics[width=0.48\textwidth]{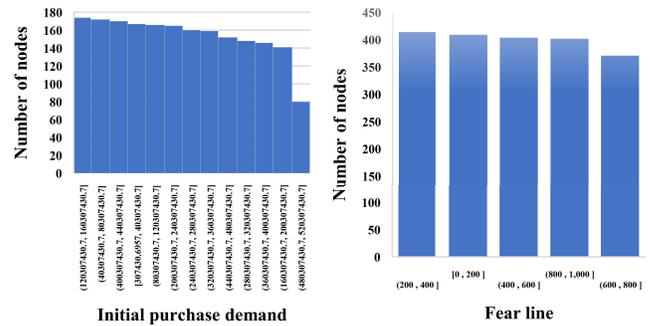}
\end{tabular}
	\caption{Node Information}
	\label{fig:img3222}
\end{figure}

We simulate \emph{20,000} nodes. They serve both as the user (the resources buyer) and the service provider (participating in generate blocks or vote for consensus). We set an amount of initial purchase demand for every node, which is presented in Figure \ref{fig:img3222}. The purchase demand of every node is randomly either increased up to $5\%$ or decreased up to $5\%$ within every ten block intervals. We set an index called "fear line" for every node. "fear line" is an indicator for if the nodes should become reliable nodes. For example, if "citizen" Bob has $5000$ currency, he uses ten currency to buy services, the time to bankruptcy is $500$ block intervals, meaning after $500$ block interval he will bankrupt if he continues buying and is not working. If Bob's "fear line" is $500$ block intervals meaning he must start to participate in the system until the time to bankruptcy is higher than $500$ again. Figure \ref{fig:img3222} shows the distribution of "fear line". When Bob submits funding to its Margin account, he ensures that he will not be bankrupted if keep buying resources during the frozen period. If the condition is fulfilled, Bob will attach a random amount of funding to the guarantee transaction.

We set $B=2$, $U=0.013$, and $AVGQ$ is the average $Q$ over 50 continuous block intervals. $50,000,000,000$ amount of currency are sent in every block iteration in the initial currency distribution at first. The amount of funding sent out decreases by $2$ in every $100$ rounds until it reaches zero. The experiment lasted $10,000$ block iterations. In the first ten block iterations, the $GPL$, $GN$ and $I$ are set to be $0.1$, $10$, $0.1$ and the $AVQ$ is $5,000,000$. The upper limit is $GPL=10,000$, $I=0.8$ while the lower limit of $I$ and $GPL$ are $0.0001$ and $10$, respectively. In this experiment, we use linear Regression of scikit-learn of python to do the linear regression.

\begin{figure}[htbp]
	\begin{tabular}{l}
   \includegraphics[trim={0cm 0 11.8cm 0},clip,width=0.48\textwidth]{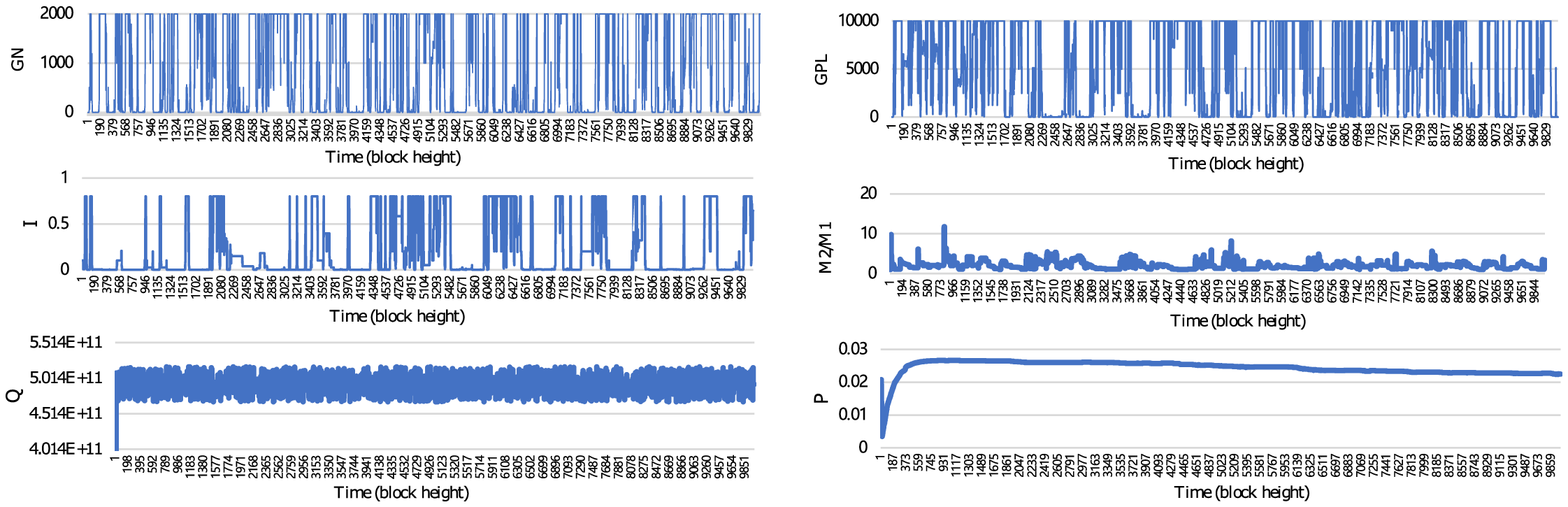}\\ \includegraphics[trim={11.5cm 0 0 0},clip,width=0.48\textwidth]{nodee.eps}
\end{tabular}
	\caption{Experiment Results}
	\label{fig:img4222}
\end{figure}

The purpose of the experiment is to show a working example of our model, the settings are arbitrary, as we cannot simulate the real usage patterns. The settings are beginning to be adjusted after the first ten block iterations. The adjustment of indexes may trigger some nodes to participate the system by achieving their "fear line". The indexes adjustment can be seen from Figure \ref{fig:img4222}. The price is very stable, and the overall purchases are also stable while the regression algorithm makes $\frac{M_2}{M_1}$ around $B$.
\section{Conclusion}
In this paper, we explored an economic model for blockchain sharding which is promising to power Grid computing. We attempted to link the price of resources with the digital labour market (the participation of nodes) and the resources purchase demand. We also attempt to stabilise and regulate the anonymous nodes by the financial mortgage. We bound the settings of Cryptocurrency with the amount of digital resources in the network. We anchor the value of cryptocurrency, in this way, the precise meaning of the cryptocurrency would be: It serves as the exchange among the computation resources worldwide.
\balance
\bibliographystyle{unsrt}
\bibliography{sample}

\end{document}